\documentclass{aa}

\usepackage{times,graphics}
\usepackage{thesisrefs}

\begin{document}

\thesaurus{11(02.08.1; 11.09.4; 11.09.1 M31; 11.19.6)}

\title{Hydrodynamic simulations of the triaxial bulge of \object{M31}}
\author{S. \, Berman}
\institute{Theoretical Physics, University of Oxford, 1 Keble Road, Oxford, England (simon@thphys.ox.ac.uk)}
\date{Received <date> / Accepted <date>}

\maketitle

\begin{abstract}
The interstellar gas flow in the inner disk of \object{M31} is
modelled using a new, two dimensional, grid based, hydrodynamics
code. The potential of the stellar bulge is derived from its surface
brightness profile. The bulge is assumed to be triaxial and rotating
in the same plane as the disk in order to explain the twisted nature
of \object{M31}'s central isophotes and the non circular gas
velocities in the inner disk. Results are compared with CO
observations and the bulge is found to be a fast rotator with a
$B$-band mass-to-light ratio, $\Upsilon_B$~=~6.5~$\pm$~0.8, and a
ratio of co-rotation radius to bulge semi-major axis,
$\mathcal{R}$~=~1.2~$\pm$~0.1, implying that any dark halo must have a
low density core in contradiction to the predictions of CDM. These
conclusions would be strengthened by further observations confirming
the model's off axis CO velocity predictions.

\keywords{hydrodynamics -- galaxies: M31 -- galaxies: ISM -- galaxies: structure}
\end{abstract}

\section{Introduction}

For many years, there has been ample evidence that bulges of barred
spiral galaxies (SB bulges) are triaxial \cite{kormendy82}. A spiral
galaxy cannot have an oblate spheroidal bulge if the position angles
of the major axes of the bulge, bar and disk are all different unless
the bulge is tipped with respect to the plane of the disk. Given that
observations of edge on galaxies show that bulges are rather flat
\shortcite{kormendy81}, it is most likely that the shortest axis is
perpendicular to the plane of the disk.

Using stellar kinematic data, \cite{kormendy82i} provide evidence that
many SA bulges (in unbarred galaxies) are rotationally flattened, and
consistent with being oblate spheroids. However, there is also
photometric and kinematic evidence to show that triaxial bulges may be
common in unbarred galaxies. In a similar way to barred galaxies, a
misalignment between the apparent major axes of the bulge and the disk
is an indicator of triaxiality, as is a twisting of the inner bulge
isophotes with respect to those of the disk. These effects have been
seen in \object{NGC 2784} \shortcite{bertola88} and in a sample of 10
unbarred galaxies \shortcite{zaritsky86}.

Stronger evidence for triaxiality comes from combining photometry with
kinematic data. In an non-axisymmetric potential, the shape of the
rotation curve will depend on the position of the line of sight and
the major axis of the non-axisymmetric component. A slowly rising
rotation curve or one in which a bump of extreme velocities is seen
near the centre are indications of triaxiality
\shortcite{gerhard89}. By comparing the theoretical gas velocity field
derived from a triaxial potential with the observed kinematic data,
one can determine whether or not the assumption of triaxiality is
justified. Triaxiality has been demonstrated for both \object{NGC
4845} \shortcite{gerhard89} and for our Galaxy \shortcite{gerhard86}.

It is important to know whether triaxiality is common in SA bulges,
since any asymmetries in the underlying potential of the inner galaxy
have a large effect on the motions of interstellar gas and could
provide a method of transporting gas into galaxy cores. Hence,
triaxial bulges could provide clues to the fueling of central
starbursts and active galactic nuclei. Furthermore, large scale
movement of gas within a spiral galaxy could lead to changes in the
overall morphology. In particular, a build up of gas in the galactic
centre can result in the creation of a bar or its destruction.

Triaxial bulges can also help to constrain the masses of dark halos in
galaxies. Kinematic observations contain the best evidence for the
existence of dark halos. Flat rotation curves imply that the mass
distribution in a galaxy must extend well beyond the observed stellar
or gaseous distributions. However, in an axisymmetric galaxy it is not
possible to determine how much of the mass of the galaxy resides in
the disk and how much in the halo. Many rotation curves can be fitted
by models ranging from zero-mass disks to `maximum disks'
\shortcite{vanalbada86}, that is disks that are as massive as possible
whilst ensuring that the halo is not hollow.

The velocity fields of barred galaxies provide extra information that
can be used to break the disk-halo degeneracy \shortcite{weiner00}
. Large jumps in velocity are an indication of triaxiality (either bar
or bulge). These velocity jumps can help to determine a mass model for
the non-axisymmetric component. Assuming a spherical halo, this can
then be used to constrain the mass-to-light ratio of the disk and the
mass fraction of the halo.

There is both photometric and kinematic evidence for triaxiality in
\object{M31}, the \object{Andromeda} galaxy. \cite{lindblad56} noted
that the apparent major axes of bulge and disk are misaligned and that
the central isophotes are twisted with respect to the isophotes of the
outer disk.  He attributed this to the presence of a triaxial bulge
rotating in the plane of the disk. \cite{stark77} showed that the
`Lindblad twist' is generated by a one parameter family of triaxial
models of the bulge of \object{M31}, where the parameter, $\phi$, is
the angle between the major axis of the bulge and the line of nodes of
the disk.

In addition, many observations of interstellar gas have detected
anomalously high velocities in both $\ion{H}{i}$ atomic gas
\shortcite{brinks84b} and CO molecular gas \shortcite{loinard95} in
the central $10'$ of \object{M31}. \cite{stark94} showed that these
velocities would occur in a simple, analytic, rotating barred
potential by calculating closed, non-intersecting, periodic orbits in
the plane of the disk and arguing that interstellar gas would largely
follow these orbits.

This paper aims to build on the work of \cite{stark77} and
\cite{stark94} in two ways. The potential of the bulge is derived from
observations, using the method of \cite{stark77} and the surface
brightness profile of \cite{walterbos88} whilst the gas flows are
modelled using a 2D, isothermal, hydrodynamics code. 

Sect.~\ref{sec-model} provides details of the construction of the
mass distribution of the bulge from the observed surface brightness
profile. Sect.~\ref{sec-hydro} describes the hydrodynamics code that
was used to determine the response of the gas in the ISM to the
background potential. Sect.~\ref{sec-res} discusses the results and
their comparison with observations. Sect.~\ref{sec-halo} comments on
the constraints on the dark halo implied by these
results. Sect.~\ref{sec-conc} concludes.

\section{Modelling \object{M31}}
\label{sec-model}

The \object{Andromeda} galaxy has been studied for over a thousand
years. It is one of the few galaxies which can be seen with the naked
eye and was mentioned by Abu I-Husain al-Sufi in his {\it Book on the
Constellations of the Fixed Stars} in 964 AD. As the closest large
spiral to our own Galaxy, \object{Andromeda} provides us with an ideal
laboratory for probing the structure and dynamics of spiral
galaxies. Due to its close proximity, high resolution photometric
observations are available of both the stellar and gaseous
components. These include maps in {\it U}, {\it V}, {\it B} and {\it
R} bands \shortcite{walterbos88}, $\ion{H}{i}$ \shortcite{brinks84b}
and CO \shortcite{loinard95}. Its large inclination angle
($77^{\circ}$) means that although it is difficult to determine
accurate positional data on its apparent minor axis, more information
is available on the kinematics of the galaxy. This and other
observational data is summarized in Table~\ref{obsparam}.

\begin{table}[htb]
\caption{Observed parameters}
\begin{tabular}{lll} 
\hline \hline
Distance of \object{M31}$^{\dag}$	& $D$ 		& 690 kpc 	\\
Total B-band absorption$^{\ddag}$	& $A_B$	& 1.3 mag	\\
Inclination angle$^{\dag}$	& $\theta$	& $77^{\circ}$ 	\\
Systemic velocity$^{\dag}$	& $v_{\rm sys}$	& $-315$ \mbox{km s$^{-1}$}	\\
Bulge effective radius$^*$	& $r_e$ 	& $10'$ 	\\
Bulge surface brightness$^*$	& $I_e$		& 22.2 mag arcsec$^{-2}$ \\
Axis ratio of inner isophotes$^*$	& $\beta$	& 1.54	\\
Bulge to disk isophote angle$^{**}$ 	& $\psi$ & $10^{\circ}$ \\
Max. bulge isophote semi-major axis$^{\S}$ & $r_{\rm max}$ & $11'.4$ \\
\hline \\
\end{tabular}
\footnotesize{\\ At the distance of \object{M31}, $1'$ = 200 pc on the major axis}\\
\footnotesize{$^{\dag}$ \cite{hodge92}, $^*$ \cite{walterbos88}}\\ 
\footnotesize{$^{\ddag}$ \cite{burstein82} and \cite{vangenderen73}}\\ 
\footnotesize{$^{**}$ \cite{lindblad56}, $^{\S}$ \cite{stark77}}
\label{obsparam}
\end{table}

The model of \object{M31} consists of two components: a rotating bulge and an
axisymmetric component combining the effects of both disk and halo.

\subsection{The bulge}

As shown in \cite{stark77}, a surface brightness profile in which the
isophotes are elliptical can be transformed into a one parameter
family of volume density profiles in which the density is constant on
similar, ellipsoidal shells. Stark's procedure relies on the use of a
number of observationally determined variables and the input of two
other parameters: the appropriate mass-to-light ratio, $\Upsilon$, and
the angle, $\phi$, between the major axis of the bulge and the disk's
line of nodes, both of which are determined by comparison with gas
kinematics.

The $B$-band surface brightness profile of the bulge of \object{M31}
follows the $r^{\frac{1}{4}}$ law of \cite{devaucouleurs58}:

\[ I(s) = I_e \, \exp(-7.67 [ (s/r_e)^{\frac{1}{4}} - 1] ) \, , \]

\noindent
where $r_e$ is the effective radius, $I_e$ is the surface brightness
at $r_e$ and $s$ is the semi-major axis length of the isophote. Table
\ref{obsparam} gives the observationally determined values for
\object{M31}. However, to recover the true bulge surface brightness
profile, this must be corrected for absorption both in \object{M31}
and in our own Galaxy. Using the values of $A_B$ =\, 0.32 mag for our
Galaxy in the direction of \object{M31} \shortcite{burstein82}, $A_B$
=\, 0.98 mag for the bulge region of \object{M31}
\shortcite{vangenderen73} and $M_{\odot B}$ =\, 5.48 for the absolute
$B$-band solar magnitude \shortcite{allen73}, the value of
$I_e$~=\,~22.2 mag arcsec$^{-2}$ in \cite{walterbos88} is transformed
into $I_e$~=\,~289~\mbox{$L_{\odot}$ pc$^{-2}$}.

To recover an ellipsoidal luminosity distribution from this brightness
profile, an Abel-type integral equation has to be solved. The solution
is itself an integral, which is numerically integrated using Romberg's
method and multiplied by a constant $B$-band mass-to-light ratio,
$\Upsilon_B$, to generate a 3D mass density profile.

The inner isophotes, which are twisted with the respect to those of
the outer disk, are used to define the spatial extent of the
bulge. Hence, the bulge model is truncated at an elliptical radius
$r_{\rm max}$~=~11$'$.4, the length of the semi-major axis of the
largest twisted isophote. For the model of minimum $\chi^2/N$, this
corresponds to a bulge semi-major axis of 3.5 kpc.

Since the hydrodynamic calculations are carried out in 2D, only the
forces in the plane of the disk are relevant. Therefore the force due
to the bulge is calculated in the plane $z$~=~0. The force at a point
$(x,y)$ is found by splitting the bulge into 300 concentric, triaxial
shells and summing the contributions of all shells interior to that
point. From \cite{hunter88}, the component of the force acting at
$(x,y)$ in the $x$ direction due to an ellipsoidal shell of semi-axes
$a_i,b_i,c_i$, where $a_i~>~b_i~>~c_i$ and $c_i$ lies perpendicular to
the plane of the disk, is

\[ dF_x = -\frac{2 \, \pi \, G \, \rho(a_i) \, b_i \, c_i \, da_i}
{\sqrt{(a_i^2 + \kappa)(b_i^2 + \kappa)(c_i^2 + \kappa)}} 
\frac{\partial \kappa}{\partial x}\, ,\]

\noindent
where $\rho(a_i)$ is the tabulated density of the shell with semi-major
axis length $a_i$ and $\kappa$ is the largest positive root of the
equation

\[ \frac{x^2}{(b_i^2 + \kappa)} + \frac{y^2}{(a_i^2 + \kappa)} = 1. \]

\noindent
A similar relationship exists for the forces in the $y$ direction.

The region inside the central shell is modelled as a homogeneous
ellipsoid of density, $\rho_c$. This central region is split into
three smaller shells and the volume weighted densities of these
smaller shells are summed to give $\rho_c$.

Table \ref{phi} details the effect that changes in $\phi$ have on the
other physical variables in the model. Larger values of $\phi$ are
associated with fatter, shorter and more rapidly tumbling bulges.

\begin{table}[htb]
\caption{Effect of changing $\phi$ on the size, shape, speed and mass of
the bulge}
\begin{tabular}{lllllllll} 
\hline
$\phi$	& $a^*$	& $b^*$ & $c^*$ & $a/b$ & $\Omega^{\dag}_{1.0}$ & $\Omega^{\dag}_{1.2}$ & $\Omega^{\dag}_{1.4}$ & $M^{\S}$\\
\hline \hline
$6^{\circ}$	& 25.0	& 12.7	& 4.80	& 2.35	& 51.2	& 44.3	& 37.4	& 0.67 \\
$12^{\circ}$	& 19.2	& 10.4	& 6.05	& 1.86	& 62.4	& 51.0	& 45.8	& 1.11 \\
$18^{\circ}$	& 16.8	& 10.1	& 6.45	& 1.66	& 70.2	& 56.2	& 49.3	& 1.35 \\
$24^{\circ}$	& 15.5	& 9.75	& 6.65	& 1.58	& 76.3	& 60.4	& 52.2	& 1.52 \\
$30^{\circ}$	& 14.6	& 9.40	& 6.80	& 1.55	& 81.3	& 63.8	& 54.7	& 1.65 \\
$36^{\circ}$	& 13.9	& 8.95	& 6.90	& 1.55	& 85.6	& 66.8	& 56.8	& 1.76 \\
$42^{\circ}$	& 13.4	& 8.45	& 7.00	& 1.58	& 89.6	& 69.5	& 58.7	& 1.85 \\
$48^{\circ}$	& 12.9	& 7.75	& 7.10	& 1.66	& 93.4	& 72.1	& 60.5	& 1.94 \\
\hline \\
\end{tabular}
\footnotesize{\\ $^*$ semi axis lengths (arcmin)}\\
\footnotesize{$^{\dag}$ bulge pattern speed (\mbox{km s$^{-1}$ kpc$^{-1}$}) for $\mathcal{R} = 1.0, 1.2$ or 1.4}\\ 
\footnotesize{$^{\S}$ bulge mass per unit mass-to-light ratio ($10^9 M_{\odot}$)}
\label{phi}
\end{table}

\subsection{The axisymmetric component}

A simple interpretation of observations of gas velocities in
\object{M31} \shortcite{brinks84b} suggests that the circular speed,
$v_c(r)$, along the line of nodes of the disk is equal to a constant
velocity, $v_0$, at radii greater than $r_0$. At radii less than
$r_0$, the circular speed falls linearly to zero. The axisymmetric
component has a mass distribution contrived to ensure that the model's
circular speed behaves thus. To determine $v_0$ and $r_0$, a two
component straight line is fitted to the CO observations of
\cite{loinard95}. The data are folded about $X$~=~0 and the four most
central points are excluded from the fit as they are fall inside the
bulge and will therefore be experiencing severe non-axisymmetric
motion. The best fit occurs for $v_0$~=~248~\mbox{km s$^{-1}$} \, and
$r_0$~=~30.0$'$. This value of $v_0$ is deprojected by a factor $\sin
\theta$ for use in the calculation of the forces due to the
axisymmetric component.

Since the bulge forces have already been calculated, the forces due to
the axisymmetric component can be found by considering that at radii
greater than $r_0$, $v_A^2 + v_B^2 = v_0^2 \sin^{-2} \theta$, where
$v_A$ and $v_B$ are the contributions to the circular speed along the
line of nodes from the axisymmetric component and the bulge,
respectively. At radii less than $r_0$, the forces due to the
axisymmetric component are assumed to fall linearly to zero.

\begin{figure}
 \resizebox{\hsize}{!}{\includegraphics{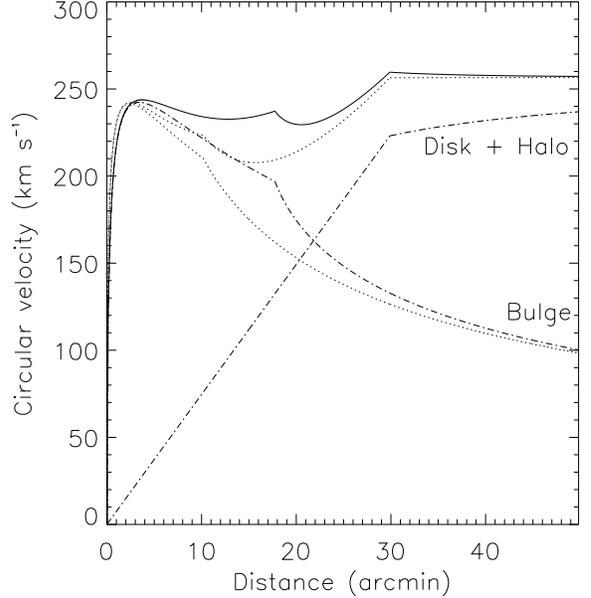}} 
 \caption{Circular speed for best fit model. The solid and dot-dashed
 lines are along the bulge major axis whilst the dotted lines are
 along the bulge minor axis}
 \label{fig:cvel}
\end{figure}

Fig.~\ref{fig:cvel} shows the circular speed curve for the model
and the contributions of both the bulge and the axisymmetric components.

Sect.~\ref{sec-halo} considers whether and how the axisymmetric
component can be broken up into seperate contributions from the disk
and dark halo.

\section{Hydrodynamic calculations}
\label{sec-hydro}

The ISM of \object{M31} is assumed to be an isothermal gas responding
to the fixed stellar potential described above. \cite{cowie80} argued
that an isothermal fluid is a reasonable assumption for a cold, dense
ensemble of clouds. The sound speed, $c_g$, is taken to be the cloud
dispersion velocity, $\sim$~10~km s$^{-1}$, and the surface density,
$\Sigma$, is the average density of the surrounding clouds.

The response of the gas is determined using a parallel, hydrodynamics
code, GALAHAD, based on the shock capturing FS2 algorithm (van Albada
and Roberts 1981, van Albada et al. 1982). The code solves the
discrete, isothermal Euler equations with gravitational source terms
on a 2D Cartesian grid, which represents half of a rotationally
symmetric galaxy. The 80 $\times$ 160 grid cells cover 10 $\times$ 20
kpc so that each grid cell is a square, 125 pc on a side. This is
approximately the same resolution as that of the CO observations
described later. The bulge is positioned so that the major axis is
aligned with the symmetry axis of the grid.

The FS2 algorithm is well suited to problems in galactic
hydrodynamics. It is a second-order, flux-splitting scheme with no
added numerical viscosity, which utilises operator splitting to
include source terms and dimensional splitting to perform calculations
in 2D. Discontinuities in the flow are handled by a limiter in the FS2
algorithm rather than by jump conditions. The self gravity of the gas
is neglected in all simulations.

The isothermal Euler equations are essentially the conservation laws
of mass and momentum in the $x$ and $y$ directions for compressible,
frictionless fluids plus an equation of state defining the pressure as
$p$~=~$\Sigma \, c_g^2$. In a frame rotating with the bulge at a pattern
speed $\Omega_p$, and with gravitational source terms, the equations
are

\[\frac{\partial \bf U}{\partial t} + \frac{\partial \bf F}{\partial
x} + \frac{\partial \bf G}{\partial y} = {\bf S + R}, \]

\[
\begin{array}{cc}

{\bf U} = 
\left( \begin{array}{c}
\Sigma \\
\Sigma u \\
\Sigma v \end{array} \right),

&

{\bf F} = \left( \begin{array}{c}
\Sigma u \\
\Sigma (u^2 + c_g^2) \\
\Sigma u v \end{array} \right), 

\end{array}
\]

\[
\begin{array}{cc}

{\bf G} = \left( \begin{array}{c}
\Sigma v \\
\Sigma u v \\
\Sigma (v^2 + c_g^2) \end{array} \right), 

&

{\bf S} = \left( \begin{array}{c}
0 \\
\Sigma (\Omega_p^2 x + 2 \Omega_p v - \Phi_x) \\
\Sigma (\Omega_p^2 y - 2 \Omega_p u - \Phi_y) \end{array} \right), 

\end{array}
\]

\noindent
where $u$ and $v$ are the gas velocities in the $x$ and $y$
directions, $\Phi_x$ and $\Phi_y$ are the gravitational potential
gradients in the $x$ and $y$ directions and {\bf R} is defined below.

The pattern speed, $\Omega_p$, fixes a co-rotation radius, $r_{\rm
co}$, at which, on the major axis, the gravitational and centrifugal
forces are equal and, in the rotating frame, gas there can stand
still. The co-rotation radius occurs at the distance along the major
axis at which $\frac{\partial \Phi_{\rm eff}}{\partial y} =$~0, where
the effective potential $\Phi_{\rm eff} = \Phi - \frac{1}{2}
\Omega_p^2 r^2$. Following \cite{athanassoula92b}, we set $r_{\rm
co}$~=~$\mathcal{R} \, a$, where $a$ is the length of the bulge
semi-major axis. However, unlike the galaxy models of
\cite{athanassoula92b}, \object{M31} is a real galaxy, modelled with a
somewhat arbitrary bulge edge. Hence, it cannot be assumed that the
value of $\mathcal{R}$ will be 1.2 $\pm$ 0.2 as quoted in
\cite{athanassoula92b}.

During every simulation, the gas gradually loses angular momentum and
falls inwards to the galactic centre. As in \cite{athanassoula92b},
GALAHAD includes a routine to compensate for this by crudely
simulating the interaction of stars with the ISM: star formation
occurs and gas is removed in areas of high density, whereas stellar
mass loss is assumed to take place at a steady rate across the
galaxy. These phenomena are modelled by setting

\[ 
{\bf R} = \alpha \left( \begin{array}{c}
\Sigma_0^2 - \Sigma^2 \\
\Sigma_0^2 u_0 - \Sigma^2 u \\
\Sigma_0^2 v_0 - \Sigma^2 v \end{array} \right), 
\]

\noindent
where $\alpha$ is the gas recycling co-efficent, $\Sigma_0$ is the
initial gas surface density and $u_0$ and $v_0$ are the initial $x$
and $y$ velocities. The values of these and the other hydrodynamical
and computational parameters used in the simulations are listed in
Table~\ref{hydroparam}.

\begin{table}[htb]
\caption{Hydrodynamical and computational parameters}
\begin{tabular}{lll} 
\hline \hline
Number of grid cells		& $I \times J$	& $80 \times 160$ 	\\
Gas recycling parameter$^{\dag}$ & $\alpha$	& 0.3 pc$^2 \, M^{-1}_{\odot}$ yr$^{-1}$ 	\\
Initial gas density		& $\Sigma_0$	& $1 \, \mbox{$M_{\odot}$ pc$^{-2}$}$		\\
Constant mass radius		& $r_M$		& 10 kpc	\\
Sound speed$^{\ddag}$		& $c_g$		& 10 \mbox{km s$^{-1}$} 		\\
Galaxy radius			& $r_G$		& 10 kpc		\\
Courant number			& $C$		& 0.5 			\\
\hline \\
\end{tabular}
\footnotesize{\\ $^{\dag}$~\cite{athanassoula92b}, $^{\ddag}$~\cite{cowie80}} 
\label{hydroparam}
\end{table}

Simulations start with the gas at a uniform surface density
$\Sigma_0$~=~1~$M_{\odot}$~pc$^{-2}$ and on circular orbits at a speed
given by $\beta v_A(r)$ with $\beta$ chosen such that $\beta^2
v_A^2(r) = G M(r_M)/r_M$, where $r_M$ = 10 kpc and $M(r_M)$ is the
mass of the model at $r_M$.

The boundary conditions are set using two rows of ghost cells, outside
the boundary of the grid, on which the dynamical variables are held
constant at their initial values.

The bulge is linearly introduced over half a rotation period and the
model is evolved to a quasi steady state over three rotation periods,
corresponding to about 0.3 Gyr. A comparison of a run at three and
four rotation periods shows very little change in the velocity fields
over much of the galaxy. However, in the region in which gas orbits
around the Lagrangian points, significant discrepancies are
found. Oscillations with amplitude $\sim$~10~km s$^{-1}$ occur in the
total gas velocity with a period of 70~Myr. These oscillations persist
until at least ten rotation periods.

A number of tests were performed to ensure that GALAHAD performs as it
should. One simple and effective test for codes used in spiral galaxy
simulations is to model an axisymmetric galaxy to determine how much
gas infall there is due to numerical viscosity. After 2~Gyr, 98\% of
the grid still had the same gas density as at the start and only 8\%
of the initial gas mass had fallen into the central 2\% of the
grid. The density in the central cells rose from
1~$M_{\odot}$~pc$^{-2}$ to 40~$M_{\odot}$~pc$^{-2}$.

\begin{figure*}
\resizebox{\hsize}{!}{\includegraphics{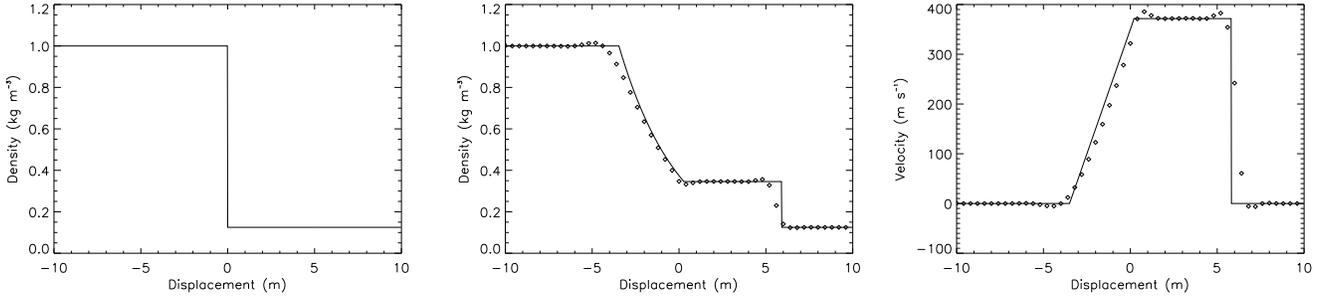}} 
\caption{1D Shock-tube test with sound speed,
 $c$~=~350~m~s$^{-1}$. The left hand panel shows the initial density
 conditions, the central panel shows the gas density after 0.01\,s and
 the right hand panel shows the gas velocity after 0.01\,s. In all
 panels, the solid line is the analytic solution and each diamond
 represents a cell in the simulated results.}
\label{fig:shocktube}
\end{figure*}

Shock-tube tests were performed on GALAHAD to test the coding of the
FS2 algorithm in the absence of source terms. A shock-tube test
involves setting up gas on either side of a thin membrane at a
constant density and at zero velocity. Model results were compared to
the analytical solution for isothermal gas in two separate scenarios:
the 1D shock-tube and the 2D oblique shock-tube. The results are
presented for the 1D test in Fig.~\ref{fig:shocktube} and show that
GALAHAD adequately replicates the analytical solutions.

\section{Simulation results}
\label{sec-res}

Overall, 87 simulation runs were performed within the three
dimensional parameter space, ($\phi$, $\Upsilon_B$, $\mathcal{R}$). The
ranges of the parameters are $12^{\circ} < \phi < 36^{\circ}$, $5.5 <
\Upsilon_B < 8.5$ and $1.0 < \mathcal{R} < 1.4$.

The model which most closely fits the observations is detailed in
Table \ref{bestfit}. The method which was used to determine this model
of minimum $\chi^2/N$ is described in Sect.~\ref{sec-obs}.

\begin{table}[htb]
\caption{The model of minimum $\chi^2/N$}
\begin{tabular}{lll} 
\hline \hline
Semi-major axis			& $a$		& 3.5 kpc 	\\
Bulge mass			& $M_{\rm bulge}$	& $8.1 \times 10^9 \, M_{\odot}$ 	\\
Bulge mass-to-light ratio	& $\Upsilon_B$	& 6.5		\\
Pattern speed			& $\Omega_p$	& 53.7 \mbox{km s$^{-1}$} kpc$^{-1}$	\\
Rotation ratio			& ${\mathcal R}$	& 1.2 		\\
Phi angle			& $\phi$	& $15^{\circ}$		\\
\hline
\label{bestfit}
\end{tabular}
\end{table}

\begin{figure*}
 \resizebox{\hsize}{!}{\includegraphics{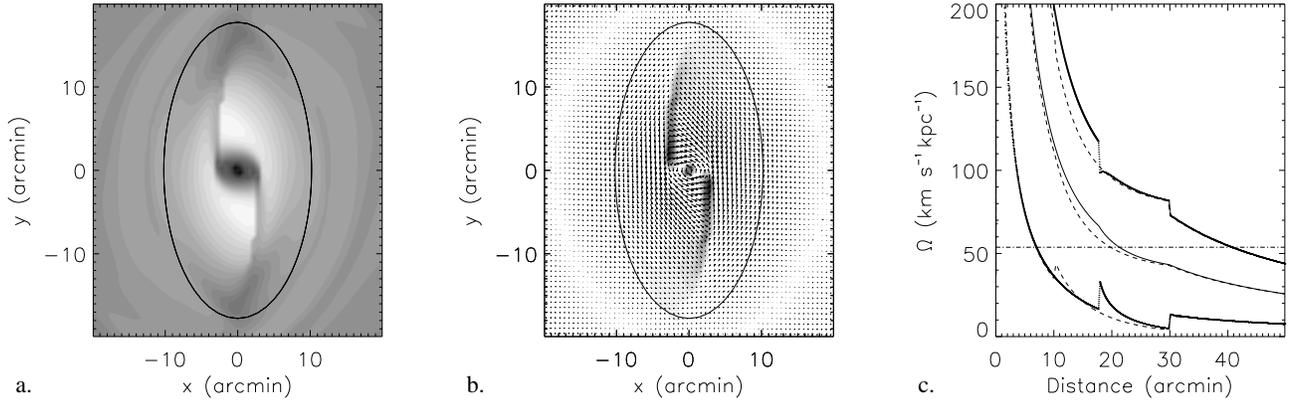}} \caption{The
 model of minimum $\chi^2/N$. Panel (a) shows as a greyscale the
 logarithmic density for the central $20'$, suggesting the existence
 of dust lanes. Panel (b) shows the velocity field for the same
 region, in the frame rotating with the bulge. The greyscale is minus
 the divergence of the velocity, again highlighting the shocks. The
 solid ellipse marks the bulge outline in both panels. Panel (c) shows
 $\Omega$ and $\Omega \pm \kappa/2$ along the line of nodes (solid
 lines) and the minor axis (dotted lines). $\Omega$ and $\kappa$ have
 been determined from numerical derivatives of the potential. The
 dot-dash line marks $\Omega_p$, the bulge pattern speed, at 54
 \mbox{km s$^{-1}$ kpc$^{-1}$}.}  \label{fig:best}
\end{figure*}

\subsection{Gas flows}

In the plane $z$~=~0 of triaxial or barred potentials, periodic orbits
exist in two main families, $x_1$ and $x_2$. $x_1$ orbits exist at all
radii and are elongated along the bulge major axis. $x_2$ orbits exist
inside the Inner Lindblad Resonance (ILR) and are oriented
perpendicular to the major axis. In the model of minimum $\chi^2/N$,
these effects can be seen in panel~(b) of Fig.~\ref{fig:best}, where
gas near the outline of the bulge is moving in orbits elongated along
the major axis.  Near the centre of the bulge, the gas flows along
orbits elongated along the minor axis.

When gas falls inwards and passes the ILR, it shifts from $x_1$ to
lower energy $x_2$ orbits. In doing so, it creates the `spray' region
seen as the diagonal gas flows just above and to the right and below
and to the left of the centre in panel~(b) of
Fig.~\ref{fig:best}. Since this area contains transient gas which is
not on any periodic orbits, it is an area of very low density and
corresponds to the low density regions in panel~(a) of
Fig.~\ref{fig:best}.

As the gas travels through the `spray' region, it slows down and
climbs out of the bulge's potential well. At the end of the `spray'
region, the gas reaches its minimum kinetic energy, abruptly changes
direction and follows the gas it has encountered back down into the
potential well. This abrupt change in velocity is seen in panel~(b) of
Fig.~\ref{fig:best} as the peak in the negative of the divergence of
the velocity, a signature of shocked gas. This corresponds to the high
density lanes of gas in panel~(a) of Fig.~\ref{fig:best}, which forms
where the shock causes gas to pile up. \cite{athanassoula92b} shows
that these shocks are intimately linked to the dust lanes which appear
in many barred galaxies.

\subsection{The dust map}

The high density gas appearing in Fig.~\ref{fig:best} panel~(a)
suggests the existence of dust lanes in \object{M31}. Using the value
for the mean $B$-band absorption in the bulge region of
$A_B$~=~0.98~mag, \shortcite{vangenderen73}, the gas density can be
calibrated and an absorption map for the bulge can be determined.

Taking the column density of interstellar hydrogen to be $N({\rm
H_{tot}}) = N({\rm H_I}) + 2 N({\rm H_2}) = $ 1.9~$\times$~10$^{25}
A_V$~m$^{-2}$~mag$^{-1}$ and $A_B$~=~1.3$ A_V$ \shortcite{bm98}, it
follows that the mean gas density in the bulge region should be
$\bar{\Sigma}_{\rm H_{tot}}$~=~11~$M_{\odot}$~pc$^{-2}$.

\begin{figure*}
 \resizebox{\hsize}{!}{\includegraphics{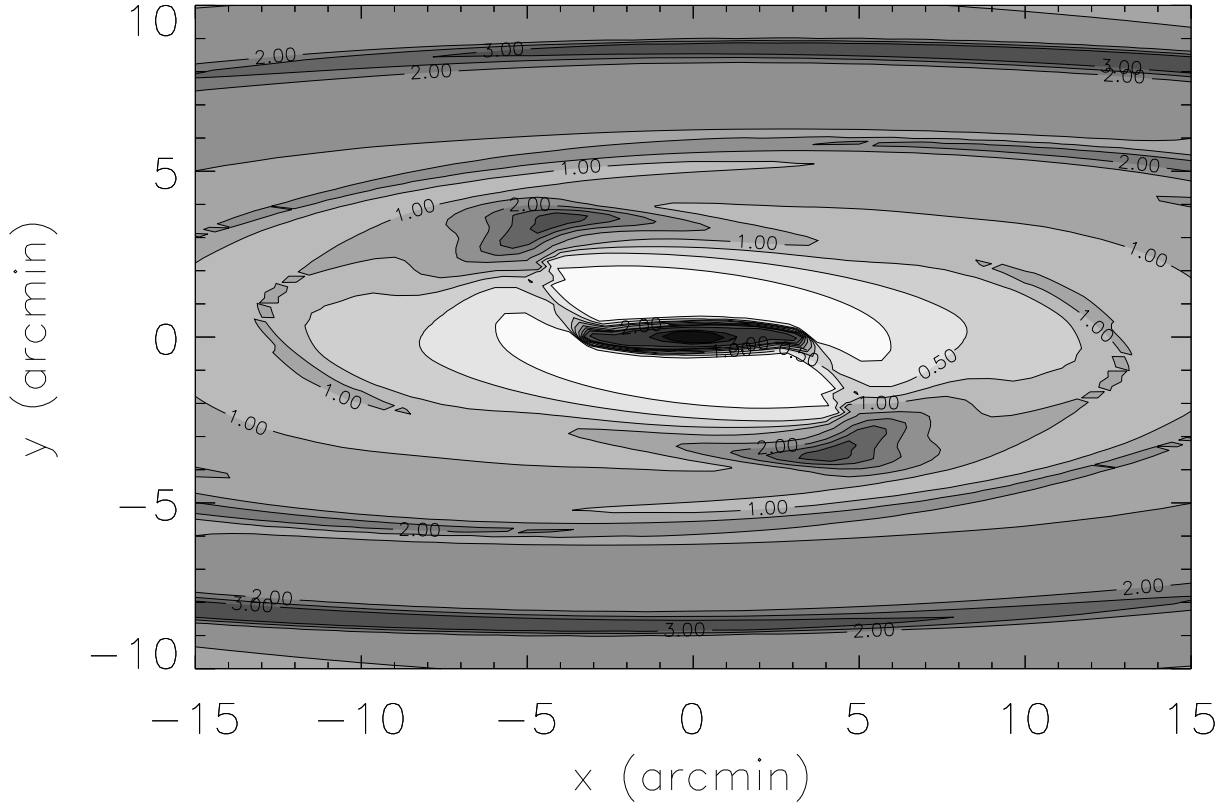}} 
 \caption{Absorption map for the bulge of \object{M31}. Derived from the
gas density field of the model of minimum $\chi^2/N$ rotated and
inclined to the observer's viewpoint. The contours and greyscale show
the {\it B}-band absorption expected from dust. Contours are at 0.25,
0.5, 0.75, 1.0, 1.5, 2.0, 2.5, 3.0, 5.0, 10, 20 and 90 {\it
B}-band magnitudes.}
 \label{fig:dust}
\end{figure*}

Assuming that CO is a reasonable tracer of both interstellar hydrogen
and dust in \object{M31} \shortcite{neininger98n}, $\bar{\Sigma}_{\rm
H_{tot}}$ can be used to calibrate the gas densities in the bulge
region and determine the corresponding {\it B}-band absorption map,
which is shown in Fig.~\ref{fig:dust}.

\subsection{Comparison of models}
\label{sec-obs}

\begin{figure*}
 \resizebox{\hsize}{!}{\includegraphics{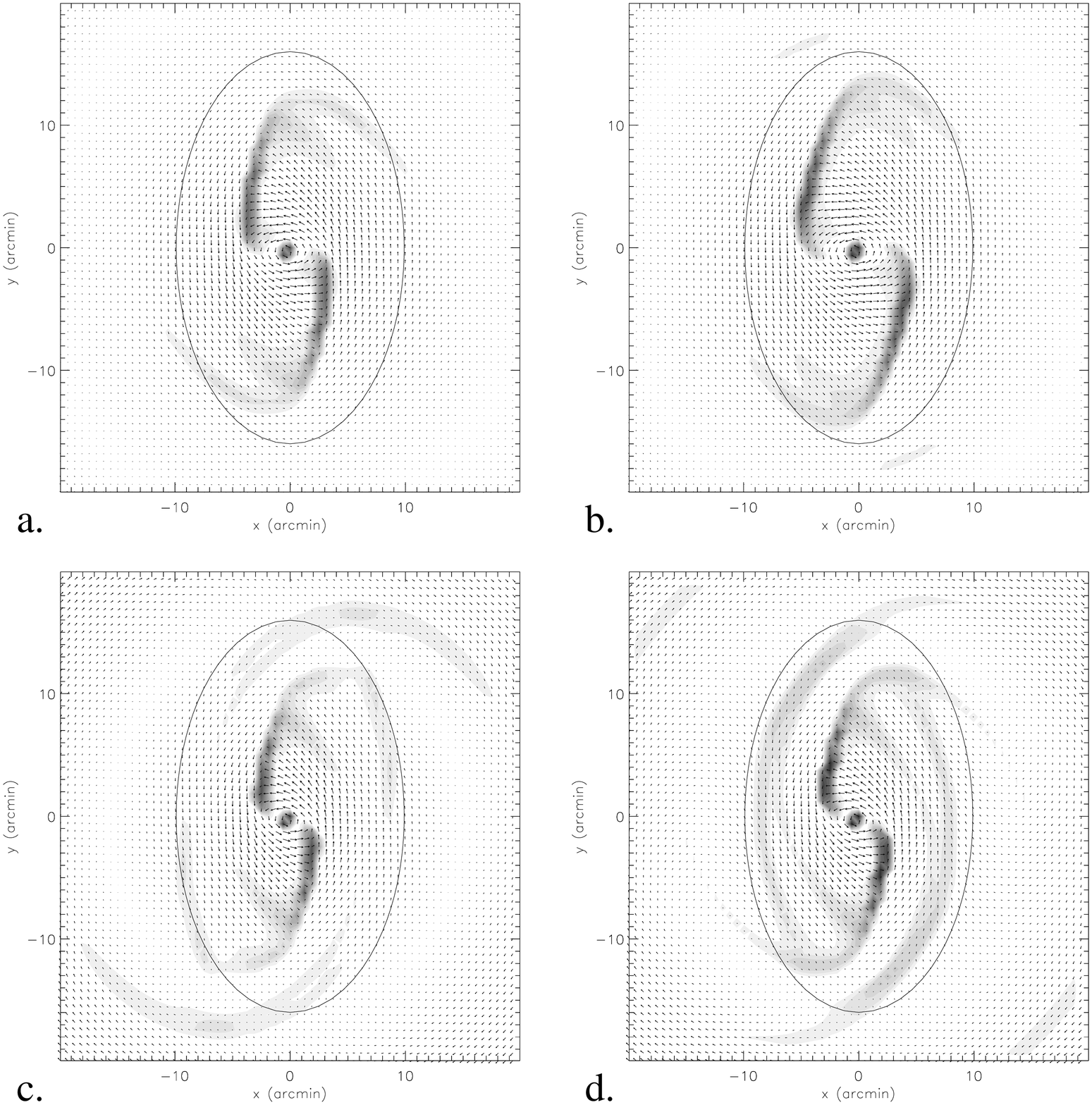}}
 \caption{Velocity fields for four different runs in the frame
 rotating with the bulge. The models in panels (a) and (b) are slower
 bulges with $\mathcal{R}$~=~1.4 whereas those in panels (c) and (d)
 are faster with $\mathcal{R}$ = 1.0. Panels (a) and (c) are lighter
 bulges with $\Upsilon_B$ = 5.5 and those in panels (b) and (d) are
 heavier with $\Upsilon_B$ = 7.5. The pattern speeds for the four
 models are $\Omega_p$ = 50, 52, 70 and 77 \mbox{km s$^{-1}$
 kpc$^{-1}$} four panels (a), (b), (c) and (d) respectively. In all
 four models, $\phi$ = 22$^{\circ}$ and the semi-major axis of the
 bulge, $a$ = 16$'$. The solid ellipse in each panel marks the bulge and
 the greyscale is minus the divergence of the velocity, highlighting
 the shocks. Every computational cell is shown.}
 \label{fig:comp}
\end{figure*}

Fig.~\ref{fig:comp} shows the velocity field for a selection of four
different models in the frame rotating with the bulge. Each model has
$\phi$~=~22$^{\circ}$ implying that the semi-major axis of the bulge
$a$~=~16$'$ for every model. The solid ellipse in each panel marks the
extent of the bulge and the greyscale is minus the divergence of the
velocity, highlighting the shocks.

The top two panels panels (a) and (b) are the slower bulges
($\mathcal{R}$ = 1.4 and $\Omega_p$ = 50 and 52 \mbox{km s$^{-1}$
kpc$^{-1}$}) whereas (c) and (d) are faster ($\mathcal{R}$ = 1.0 and
$\Omega_p$ = 70 and 77 \mbox{km s$^{-1}$ kpc$^{-1}$}). The left hand
panels, (a) and (c), describe lighter bulges ($\Upsilon_B$ = 5.5) and
the two on the right, panels (b) and (d) describe heavier bulges
($\Upsilon_B$ = 7.5).

A comparison of the top two panels with the bottom two shows the
effect of a faster bulge on the shock morphology. A faster bulge has a
smaller co-rotation radius and radius of the ILR at which gas
transfers between $x_1$ and $x_2$ orbits. Since this transfer of gas
between families of periodic orbits is the primary cause of the shocks
seen in Fig.~\ref{fig:comp}, we see that fast bulges have shocks
which occur closer to their centres. This is the case irrespective of
whether the bulge is heavy or light. The faster bulges also have
pronounced secondary shocks, which emerge near the end of the major
axis of the bulge.

The left hand panels (a) and (c) show the lighter bulges. In comparison
with the heavier bulges of panels (b) and (d), they have weaker
shocks. The shocks for the lighter bulges are also slightly closer to
the centre and slightly straighter than for the heavier ones.

\subsection{Comparison with observations}

The observational data used to judge the success of the simulations is
CO data from \cite{loinard95}. The molecular gas is likely to be
confined to the plane of the disk and follow the periodic orbits
described above. CO is a good tracer of the majority of molecular gas
in a galaxy, particularly in the case of \object{Andromeda}
\shortcite{neininger98n}. Hence, it is amenable to modelling using
isothermal hydrodynamics.

However, although there have been recent CO maps made of most of the
\object{Andromeda} galaxy (Loinard et al. 1999, Dame et al. 1993),
there is very little CO data available in the central bulge
region. The reason for this deficiency in data could be that the CO is
very cold and difficult to detect in the inner disk
\shortcite{loinard98} or simply not there \shortcite{melchior00}.

The data from \cite{loinard95} were taken using the 30-m IRAM
millimeter radio telescope in 1993. 26 positions along the apparent
major axis of \object{M31} were surveyed and CO was detected in 16 of
them with a resolution of 35$'' \times$ 2.6~km~s$^{-1}$. All 16 data
points are used in the comparison with the models.

\begin{figure}
\resizebox{\hsize}{!}{\includegraphics{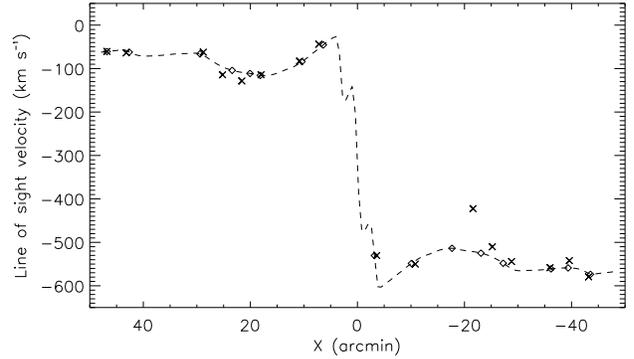}} 
\caption{Gas velocity from the model of minimum $\chi^2/N$ on the line
of nodes of \object{M31}. Crosses are CO detections. The dashed line
is data from the model on the line of nodes after transforming to the
observing frame. Diamonds are the closest points to each data point on
the dashed line in terms of data error bars. The coordinate system is
that of Baade and Arp (1964).}
\label{fig:coobs}
\end{figure}

Each model was rotated by $\phi$ and inclined by $\theta$ to match the
viewing position on the sky. A cut was then taken along the line
of nodes of the disk, involving an interpolation from off-axis positions,
and a comparison made with the CO data. Fig.~\ref{fig:coobs} shows a
comparison between the output of the best fit model and the CO data.

The model fits the data quite well except for one point at
$X$~=~-21.6$''$, $v_{\rm los}$~=~-422.4~km~s$^{-1}$ which cannot be
fit using the model described above. This anomalous point may be due
to a molecular cloud orbiting above the plane of the disk. In this
case, its line of sight velocity would be smaller than if it moved in
the plane, since it would be partly moving in the $z$ direction.

To make a quantitative statement about how good a fit a particular
model is and which model gives the best fit to the data, a likelihood
function is derived, where the errors in the position and velocity of
the CO data are assumed to be Gaussian. Given that there are errors in
both $X$ and $v_{\rm los}$, the difference between the data and the
model is found by comparing each datum with the point in the model
which is closest to it. This takes place within error space, where all
distances are divided by the error on $X (35'')$ and all velocities by
the error on $v_{\rm los}$ (2.6 \mbox{km s$^{-1}$}).

This procedure gives a value of $\chi^2/N$ for each model and a simple
method of ascertaining the model of minimum $\chi^2/N$. For this
model, $\phi$~=~15$^{\circ}$, $\Upsilon_B$~=~6.5, $\mathcal{R}$~=~1.2
and $\chi^2/N$ = 4.7 or 2.7 with or without the far flung data point
respectively. Given the various assumptions that have gone into this
model, it is unlikely that values much closer to $\chi^2/N$~=~1 could
be attained. Table~\ref{bestfit} provides details of the parameters of
this model.

\begin{figure*}
 \resizebox{\hsize}{!}{\includegraphics{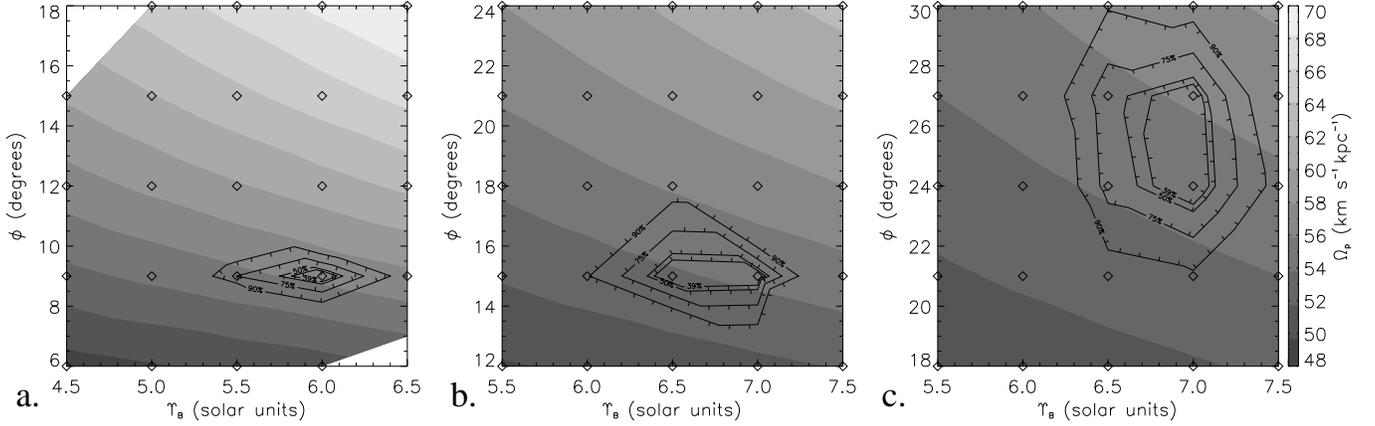}} 
 \caption{$\chi^2/N$ plots for $\mathcal{R}$ = 1.0, 1.2 and 1.4 in
panels (a), (b) and (c) respectively. The spurious point has been
removed from the data sets. Diamonds mark the position of each
model. Contours show the 39\%, 50\%, 75\% and 90\% confidence levels,
whilst the greyscale represents the bulge pattern speeds, $\Omega_p$,
for each model. The value of the best fit model in each set is
$\chi^2/N$ = 2.9, 2.7 and 3.3 for $\mathcal{R}$ = 1.0, 1.2 and 1.4
respectively.}
 \label{fig:cont}
\end{figure*}

Contour plots of $\chi^2/N$ are shown in Fig.~\ref{fig:cont} for
$\mathcal{R}$~=~1.0, 1.2 and 1.4. In each plot, the value for the
bulge mass-to-light ratio is constrained to 0.5 either side of the
best fit model value at the 90\% confidence level. However, as the
value of $\mathcal{R}$ rises from 1.0 to 1.4, $\phi$ becomes less well
constrained. This effect happens concurrently with a narrower spread
of pattern speeds. The result is that the pattern speed varies by 3 or
4~km~s$^{-1}$~kpc$^{-1}$ in each of the panels in Fig.~\ref{fig:cont}
in the 90\% confidence regions.

\begin{figure}
 \resizebox{\hsize}{!}{\includegraphics{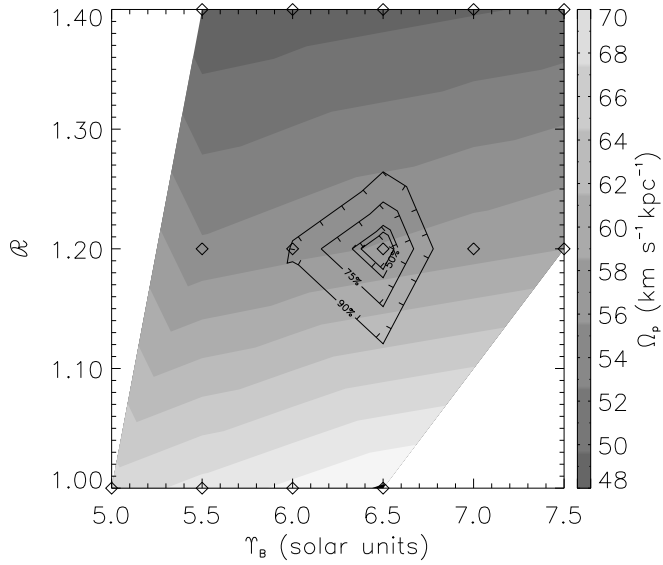}}
 \caption{$\chi^2/N$ plot for $\phi$~=~18$^{\circ}$ using the data set
 without the anomalous point. Diamonds mark the position of each
 model. Contours show the 39\%, 50\%, 75\% and 90\% confidence levels,
 whilst the greyscale represents the bulge pattern speeds, $\Omega_p$
 of each model. The value of the best fit model is $\chi^2/N$ = 3.1
 for $\mathcal{R}$ = 1.2 and $\Upsilon_B$ = 6.5.}  
\label{fig:contphi}
\end{figure}

Fig.~\ref{fig:contphi} shows the value of $\chi^2/N$ as a function of
$\Upsilon_B$ and $\mathcal{R}$ at a constant value of
$\phi$~=~18$^{\circ}$. Although the plot was only constructed at a
single value of $\phi$, it does indicate that the favoured value of
$\mathcal{R}$ lies between 1.1 and 1.3. This shows that the bulge of
\object{Andromeda} is a fast rotator.

Combining all of the above results implies that, at the 90\%
confidence level, $11^{\circ} < \phi < 24^{\circ}$, $5.7 < \Upsilon_B
< 7.3$ and $51 < \Omega_p < 55 \, \mbox{km s$^{-1}$ kpc$^{-1}$}$.

\subsection{Off axis predictions}

Unfortunately, the derived values of $\chi^2 / N$ are not very
robust. Using the above method, it is possible to vary the value of
$\chi^2 / N$ by two simply by taking the result of a particular model
at four rather than at three bar rotations. This is enough to turn a
model which fits the data very well into one which is a much worse
fit. Therefore it becomes much more difficult to determine whether a
particular model is a good representation of \object{M31} or not and
how well the above results can be trusted.

\begin{figure}
 \resizebox{\hsize}{!}{\includegraphics{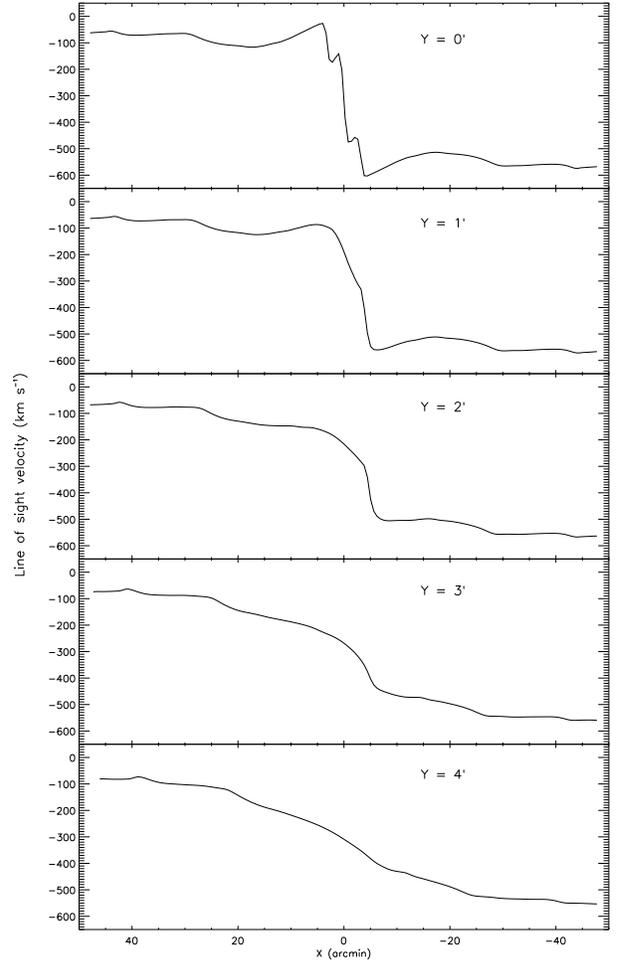}} 
\caption{Gas velocities from the model of minimum $\chi^2/N$. Cuts are
 taken at 0$'$, 1$'$, 2$'$, 3$'$ and 4$'$ parallel to the line of
 nodes of \object{M31}. The coordinate system is that of Baade and Arp
 (1964).}
\label{fig:cooff}
\end{figure}

A much stronger case could be made if more off axis CO data was
available in the inner regions of \object{M31}. In order to anticipate
such observations, I have provided in Fig.~\ref{fig:cooff} a set of
predictions from the model of minimum $\chi^2/N$ for off axis gas
velocities, for comparison with observations. All predictions are for
cuts parallel to the major axis, taken at 1$'$ intervals from 0$'$ to
4$'$ inclusive.

These results were all calculated using a distance to \object{M31} of
690 kpc for ease of comparison with previous work. However, using
recent results from Hipparcos, \cite{stanek98} deduce that the
distance is 784 kpc. Adopting this value means that
1$'$~=~230~pc. Lengths, whether observed or deprojected and the mass
of the bulge, all scale in proportion to the distance whereas the
central density scales as the inverse square of the distance to
\object{M31}.

\section{The dark halo}
\label{sec-halo}

As mentioned above, the axisymmetric component is assumed to comprise
a disk and any halo which exists in \object{M31}. This was done to
reduce the number of free parameters in the model whilst remaining
true to the observations. However, it is extremely important to
determine what fraction by mass of the inner region of \object{M31}
consists of dark matter. The inner region of \object{M31} is defined
here as a sphere inscribed by the semi-major axis of the bulge of the
minimum $\chi^2/N$ model i.e. a sphere of radius 3.5 kpc.

In order to do this, the total circular speed curve of the model of
minimum $\chi^2/N$, shown in Fig.~\ref{fig:cvel}, is taken to
represent the combined mass distribution of the bulge, disk and
halo. Hence, the total mass at 3.5~kpc, $M_T$~=~2.2~$\times$~10~$^{10}
M_{\odot}$. The contribution of the bulge to this has already been
deduced to be 8.1$~\times$~10$^9 M_{\odot}$, however there remains a
degeneracy between the contributions of the disk and the halo.

\subsection{Predictions of CDM for the dark halo}

One method of breaking this degeneracy is to represent the halo by an
NFW profile \shortcite{navarro96} derived from cosmological $N$-body
simulations,

\[ \frac{\rho(r)}{\rho_s} = \frac{1}{(r/r_s)(1 + r/r_s)^2} ,\]

\noindent
where $\rho_s$ is a characteristic density, $r_s$~=~$r_{\rm 200} / c$
is a characteristic radius and $c$ is a dimensionless concentration
parameter. $r_{\rm 200}$ is the radius within which the mean density
of the halo is 200 $\rho_{\rm crit}$, where $\rho_{\rm crit} = 3 H^2 /
8 \pi G$ is the critical density of the universe, $H$~=~$h \, H_0$ is
Hubble's constant, $H_0$~=~100~\mbox{km s$^{-1}$ Mpc$^{-1}$} and $h$
is taken to be 0.75.

The circular velocity curve for an NFW profile is given by

\[ \left(\frac{v_{\rm halo}(r)}{v_{\rm 200}}\right)^2 = \frac{1}{x} \frac{{\rm ln} \, (1 + cx) - (cx)/(1+cx)}{{\rm ln} \, (1+c) - c/(1+c)} ,\]

\noindent
where $v_{\rm 200}$ is the velocity at $r_{\rm 200}$ and $x$~=~$r/r_{\rm
200}$. From \cite{navarro97}, at a redshift $z$~=~0,

\[ r_{\rm 200} = \left(\frac{h \, v_{\rm 200}}{\rm \mbox{km s$^{-1}$}} \right) {\rm kpc} .\]

\noindent
Hence, for a given value of Hubble's constant, the halo circular speed
curve is fully specified by $v_{\rm 200}$ and $c$. \cite{navarro97}
find a relationship between the characteristic density and radius (or,
equivalently, between $c$ and $v_{\rm 200}$) of a halo in a given
cosmology, thereby reducing the NFW profiles to a one parameter
family. 

If $v_{\rm 200}$~=~$v_0$~=~257~\mbox{km s$^{-1}$}, then the method of
\cite{navarro97} gives a halo mass fraction at 3.5 kpc of 58\% and a
disk mass of only 5.8~$\times$~10$^9 M_{\odot}$. However,
\cite{navarro96} state that the maximum rotation velocity can be as
high as $1.4 \, v_{\rm 200}$. If $v_{\rm 200}$~=~$v_0 /
1.4$~=~184~\mbox{km s$^{-1}$}, the halo mass fraction at 3.5 kpc falls
to 44\% and the disk mass rises to 7.8~$\times$~10$^9 M_{\odot}$. But
in either case, the inner region of \object{M31} would be heavily dark
matter dominated if NFW halo profiles are to be believed.

\subsection{Fast bars mean minimal halos}

\cite{debattista00} convincingly demonstrate that fast, long lived
bars cannot co-exist with massive dark halos. Their argument is based
on the idea that dynamical friction from a dense, dark matter halo
slows down any rapidly rotating bar. Hence, any old bar which is found
to be a fast rotator precludes the existence of a dark halo, other
than one with the lowest possible central density consistent with not
being hollow.

`Fast' bars have $1.0 < \mathcal{R} < 1.4$ and, of the few galaxies
where $\mathcal{R}$ has been measured, all have fallen within this
range implying that all barred galaxies have minimal halos and, hence,
maximum disks. Further, unless the distribution of halo densities is
bimodal, this conclusion of minimum halos should be extended to all
spiral galaxies.

Since the model of minimum $\chi^2/N$ for \object{M31} has
$\mathcal{R}$~=~1.2 and the bulge of M31 is older than 6 Gyr
\shortcite{trager00}, \object{M31} must have a low density dark matter
halo and a maximum disk. The exact value of the halo mass fraction
inside 3.5 kpc is dependent on the profile adopted for disk and halo,
even for a maximum disk. The maximum disk model of \cite{kent89}
implies a halo mass fraction of just 0.5\% at 3.5 kpc and, although
the model of minimum $\chi^2/N$ is unable to make such a quantitative
prediction, it is clearly at odds with the NFW halo predicted by CDM.

\section{Conclusions}
\label{sec-conc}

The mass distribution for the bulge of \object{M31} has been
determined from observations of the bulge surface brightness
profile. The bulge is triaxial and rotates about the same axis as the
disk. The GALAHAD code, based on the FS2 algorithm, has been used to
reproduce observations of CO position and velocity in the bulge of the
unbarred \object{Andromeda} galaxy, particularly the extreme
velocities which appear in the central $10'$.

By matching the model gas velocity field to the observations, it has
been possible to constrain the $B$-band bulge mass-to-light ratio,
$\Upsilon_B$, and the bulge pattern speed, $\Omega_p$, at the 90\%
confidence level, to $5.7 < \Upsilon_B < 7.3$ and 51 $< \Omega_p <$ 55
\mbox{km s$^{-1}$ kpc$^{-1}$}. The angle between the major axis of the
bulge and the line of nodes of the disk has been constrained to
$11^{\circ} < \phi < 24^{\circ}$. For the model of minimum $\chi^2/N$,
$\Upsilon_B$~=~6.5, ${\mathcal R}$~=~1.2, the semi-major axis
$a$~=~3.5~kpc and $\phi$~=~15$^{\circ}$.

The procedure described above improves on previous bulge models by
constructing the mass distribution from observations. Further, the use
of a hydrodynamics code to determine the response of the gas to the
underlying potential provides a more accurate representation of the
gas velocity field than calculations of periodic orbits. However, the
oscillations inherent in a hydrodynamic model of a galaxy limit the
accuracy of any simulation and may lead to more slack in the
predictions than the formal errors imply.

If the bulge really is triaxial, it should be possible to constrain
its parameters further by attempting to observe the offset dust lanes
that should accompany the gas shocks described above. An absorption
map has been provided to assist dust lane observations.

Since the bulge of \object{M31} has been shown to be a fast rotator,
it follows from \cite{debattista00} that \object{M31} has a maximum
disk and a halo mass fraction within 3.5 kpc of just a few
percent. This is in stark contrast to the predictions of
\cite{navarro96}, which imply that the halo mass fraction within 3.5~
kpc is between 44\% and 58\%.

These conclusions would be strengthened by the confirmation of the off
axis gas velocity predictions which have been made above. This could
be done by confronting the model with further CO observations of the
central 15$'$ of \object{M31}.

Given that triaxiality has now been demonstrated in both our Galaxy
and our nearest large neighbour, \object{M31}, it seems likely that
triaxial bulges are a common feature of both SA and SB spiral
galaxies.

\begin{acknowledgements}
I thank James Binney for many useful discussions and helpful advice,
Julio Navarro for the NFW subroutine and the Oxford Supercomputing
Centre for the use of OSCAR, an SGI Origin 2000 parallel computer, to
carry out the hydrodynamic calculations.
\end{acknowledgements}

\end{document}